\documentclass{article}
\usepackage{mlspconf}
\usepackage{amssymb,amsmath,graphicx,epstopdf,booktabs,subfigure,multirow,array,setspace,,algorithm,algpseudocode,adjustbox,pbox}
\usepackage{textcomp}

\DeclareMathOperator*{\argmin}{argmin}

\toappear{Preprint: 2015}

\title{Making Sense of Randomness: An Approach For Fast Recovery of Compressively Sensed Signals}
\name{ V. Abrol and P. Sharma and A. K Sao}
\address{School of Electrical and Computing Engineering,
Indian Institute of Technology, Mandi, India\\
  {\small \tt vinayak\_abrol@students.iitmandi.ac.in, pulkit\_s@students.iitmandi.ac.in, anil@iitmandi.ac.in.}}

\begin{document}
%

\maketitle
\begin{abstract}
In compressed sensing (CS) framework, a signal is sampled below Nyquist rate, and the acquired compressed samples are generally random in nature. However, for efficient estimation of the actual signal, the sensing matrix must preserve the relative distances among the acquired compressed samples. Provided this condition is fulfilled, we show that CS samples will preserve the envelope of the actual signal even at different compression ratios. Exploiting this envelope preserving property of CS samples, we propose a new fast dictionary learning (DL) algorithm which is able to extract prototype signals from compressive samples for efficient sparse representation and recovery of signals. These prototype signals are orthogonal intrinsic mode functions (IMFs) extracted using empirical mode decomposition (EMD), which is one of the popular methods to capture the envelope of a signal. The extracted IMFs are used to build the dictionary without even comprehending the original signal or the sensing matrix. Moreover, one can build the dictionary on-line as new CS samples are available. In particularly, to recover first $L$ signals ($\in\mathbb{R}^n$) at the decoder, one can build the dictionary in just $\mathcal{O}(nL\log n)$ operations, that is far less as compared to existing approaches. The efficiency of the proposed approach is demonstrated experimentally for recovery of speech signals.
\end{abstract}
\begin{keywords}
Speech Processing, Compressed Sensing, dictionary learning, empirical mode decomposition.
\end{keywords}
\section{Introduction}
Compressed sensing (CS) or sparse signal representations have recently drawn much interest in the field of speech processing e.g., speech encryption \cite{ES84} and speech recognition \cite{ES83}. In particular, CS enables us to reconstruct a signal $\mathbf{x} \in \mathbb{R}^{n}$ which can be sparsely represented in an overcomplete dictionary $\mathbf{\Psi} \in \mathbb{R}^{n\times d}$ ($d$=$n$ for complete dictionary), via recovery of its sparse representation $\mathbf{a} \in \mathbb{R}^{d}$ from very few measurements $\mathbf{y} \in \mathbb{R}^{m}$ sampled using a measurement matrix $\mathbf{\Phi} \in \mathbb{R}^{m\times n}$ with $\it {m\ll n}$~\cite{ES15} \cite{ES18}. CS measurements are robust to degradations such as random perturbations or noise and does not require much memory for storage or transmission \cite{ES15}. In CS, although the signal acquisition is random, the obtained linear projections or measurements still preserve the relative distance between two signal points \cite{ES15}. This was supported by our observation that the compressive samples indeed preserves the envelope of the actual signal. It is a known fact that in case of speech signals, the signal envelope is very important in perception, e.g., the words are identified according to their envelope \cite{ES131}. Thus, this paper essentially focuses on speech signals.

Exploiting the envelope preserving property of CS measurements, we propose a novel method where the aim is to express a speech signal as a sparse linear combination of prototype signals extracted from  compressive speech samples directly. These prototype signals, can be intrinsic mode functions (IMFs) extracted using empirical mode decomposition (EMD), which is one of the popular methods to capture the envelope of a signal. We show that the IMFs extracted from compressive speech show similar behavior to the ones extracted from the speech signal directly. Hence, the extracted IMFs can be used to build the dictionary, using which one can recover the original speech signal from CS samples.

\subsection{Related Works}
The estimation of sparse vector (or equivalently the original signal) using compressive samples is very much influenced by the choice of dictionary \cite{ES15}. It has been shown that a sparse representation, estimated using a learned dictionary as compared to an analytic dictionary (e.g., DCT), results in better recovery of the signal\cite{ES18}. Thus, the DL problem aims to find a dictionary $\mathbf{\Psi}$ such that the error, $\Vert \mathbf{x}_i-\mathbf{\Psi a}_i \Vert^2_{2} \; \forall_i$  is minimized and $\mathbf{a}_i$ is sparsest \cite{ES17}. Typically this is achieved by alternating minimization over $\mathbf{a}_i$'s and $\mathbf{\Psi}$, i.e., the optimization is realized over one, keeping the other fixed \cite{ES18}. Details of various dictionary algorithms can be found in \cite{ES129}. Provided the dictionary is available, one can efficiently recover a speech signal from compressive speech samples via recovery of its sparse representation \cite{ES82}. For instance, approaches in \cite{ES12} and \cite{ES19}, recover a speech signal using a dictionary build from the pre-estimated vocal tract filter coefficients or line spectral frequency (LSF) code book derived from the training data. However, when only compressive samples are available, recovering the actual signal while simultaneously learning a dictionary is a difficult task. To address this issue, recent works have proposed some modified DL methods (e.g., partial-KSVD \cite{ES130}) where the dictionary is learned from CS samples by minimizing the objective function $\Vert \mathbf{y}_i-\mathbf{\Phi\Psi a}_i \Vert^2_{2} \; \forall_i$. However, such DL methods are computationally expensive, and assume that the signal support set (non-zero index locations of sparse vector) is known \textit{a priori}. Alternatively, one can use recovery based DL methods, that are mathematically tractable compared to conventional methods \cite{ES82,ES18}. Here, with an initial dictionary, the current estimate of the recovered signal from compressive samples is used to update the dictionary, and this procedure is performed iteratively until convergence. Recovery based DL methods are essentially based on the concepts of blind compressed sensing \cite{ES117}. One such iterative DL approach for speech signals is presented in \cite{ES28}.
 
Nevertheless, applying CS on speech signals involve two main issues: (1) for speech signals (which has lot of variations due to speaker, speaking style or spoken language) the dictionary should preferably be trained on speaker specific training data, which might not be available in each scenario and requires a huge amount of storage, (2) existing recovery based or conventional DL algorithms have large computational complexity. 

\subsection{Contributions of the Proposed Work}
In this paper, we propose a novel fast unsupervised DL approach for recovery of compressive speech signals. As far as this work is concerned, we are interested in the scenario where only compressed measurements of the actual speech signal are available with out any prior knowledge of signal's support set. We show that it is indeed possible to learn a dictionary from compressive speech samples, by bypassing the reconstruction of actual speech signal i.e., eliminating the abundant cost of recovering irrelevant data. To this aim, the dictionary is build using IMFs extracted directly from CS samples, without even comprehending the original speech signal or the sensing matrix used to acquire the signal. Moreover, the extracted IMFs being orthogonal results in a dictionary having good mutual coherence properties. It is worth emphasizing that the goal of the paper is not to outperform a state-of-the-art CS recovery method but is to propose an approach which can perform with an acceptable level of accuracy in heavily resource-constrained environments, both in terms of storage and computation. To the best of our knowledge, none of the previous papers have proposed such methods for compressively sensed signals. 

The rest of the paper is organized as follows: In Section~\ref{sec:1}, we briefly explains the modeling of speech signals using CS framework, and how envelope of a speech signal is preserved in compressive samples. In Section~\ref{CSEMD} we propose an efficient DL algorithm for compressive speech signals using EMD, and the experimental results are shown in Section~\ref{EXP}. The summary of paper is given in Section~\ref{sec:summary}.

\section{Modeling Speech Signals using CS}
\label{sec:1}

In CS framework, signals are sampled at less than the Nyquist rate \cite{ES82}. In particular, given a matrix $\mathbf{Y} \in \mathbb{R}^{m\times l}$ consisting of $L$ compressive speech signal frames $\{\mathbf{y}_i\}_{i=1}^L$ as columns, the recovery of the corresponding signal set ($\mathbf{X} \in \mathbb{R}^{n\times l}$) is formulated as:
\begin{eqnarray}
\small
\begin{aligned}
\mathbf{\hat{X}}&\approx\mathbf{\Psi\hat{A}}\;\; \mathrm{where}\; \mathbf{\hat{A}}\; \mathrm{is\;computed\;as,} \\ 
\mathbf{\hat{A}}=\argmin_\mathbf{A}\; & f(\mathbf{A})\; \colon   \Vert \mathbf{Y}-\mathbf{\Phi\Psi}\mathbf{A}\Vert_F^2=\Vert \mathbf{Y}-\mathbf{D}\mathbf{A}\Vert_F^2 < \epsilon,
\end{aligned}
\label{eq1} 
\end{eqnarray}
where $\mathbf{A} \in \mathbb{R}^{d \times l} $ is the sparse coefficient matrix corresponding to $\mathbf{X}$, $\epsilon$ is the error tolerance, $f()$ is a function (e.g., $l_1$-norm) that promotes sparsity and $\mathbf{D} \in \mathbb{R}^{m \times d}$ is the overall dictionary. According to CS theory, if the matrix $\mathbf{\Phi}$ satisfies restricted isometry property (RIP), and is incoherent with the dictionary $\mathbf{\Psi}$, the signal can be recovered with very high probability by linear programming methods \cite{ES15}.

\subsection{Randomness Do Make Sense: Properties of Compressive Samples}
\begin{figure}[t!]
\centering
\adjincludegraphics[trim={0 0 0 0},clip,width=6cm,height=5.3cm]{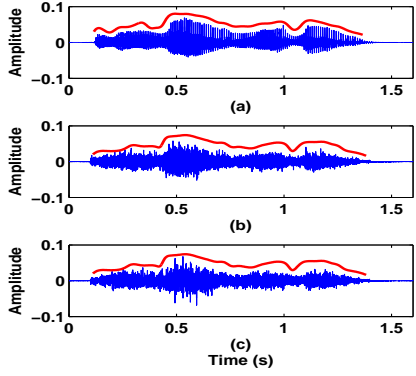}
\caption{\small Comparison of the envelopes (manually marked red) of (a) original speech signal,  (b) and (c) interpolated compressive speech signal orignally sampled at compression ratio ($m/n$) of $0.7$ and $0.5$ respectively.}
\label{fig1}
\end{figure}

CS acquires random signal measurements,\footnote{The elements of the sensing matrix are assumed to be i.i.d. random variables} and hence do not preserve any signal structures in their raw form. However, these linear projection acquired using $\mathbf{\Phi}$, which satisfies the RIP property, still preserves the relative distance between two signal points or vectors \cite{ES15}, i.e.,: 
\begin{equation}
\Vert\mathbf{\Phi}(\mathbf{x}_1-\mathbf{x}_2)\Vert_2^2\approx\Vert\mathbf{x}_1-\mathbf{x}_2\Vert_2^2\; \forall\; \mathbf{x}_1,\mathbf{x}_2 \in \mathbb{R}^n
\end{equation} 
Moreover, the mean of measured energy is exactly equal to $\Vert\mathbf{x}\Vert_2^2$ i.e., $\mathbf{E}\left[\Vert\mathbf{\Phi x}\Vert_2^2\right]=\Vert\mathbf{x}\Vert_2^2$. To illustrate this, Fig. \ref{fig1}, shows a example of the original and compressively sensed speech signal. Note that the sampling rate of compressive speech is less than that of the original speech signal, and for a fair comparison, the interpolated compressive speech, computed using cosine interpolation is plotted in the figure. It can be observed that though the measurement vector exhibits some random noise-like nature, envelopes of both the original and the compressive speech signal are approximately similar, even at different compression ratios. In other words, the preserved structure of the instance space i.e., speech signal is more prominent if viewed globally or in longer windows. 

To exploit this preserved envelope, one may decompose a compressive speech signal to extract prototype signals to build the dictionary. One way to achieve this is to apply EMD on compressive speech signal. EMD exploits the signal envelope or evolution of a signal between two consecutive local extrema to decompose a signal into orthogonal modes or IMFs, which can be used as dictionary atoms.
\begin{figure}[t!]
\centering
\subfigure[]{\adjincludegraphics[trim={0 20 0 25},clip,scale=0.4]{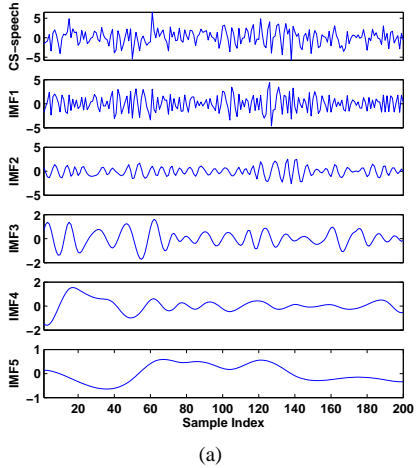}}
\caption{\small EMD decomposition of a voiced frame of compressive speech sampled at compression ratio ($m/n$) of $0.5$}
\label{fig2}
\end{figure}
\section{CS-EMD: A Fast Dictionary Learning Approach for Compressively Sensed Speech Signals}
\label{CSEMD}
The proposed approach is a exemplar based approach where a speech frame is sparsely represented as a linear combination of few IMFs from the dictionary, selected optimally using sparsity constraints. However, the IMFs used to build the dictionary are extracted directly from CS samples. Using the EMD method a given compressive speech frame $\mathbf{y}$ can be expressed as
\begin{equation}
\mathbf{y}=\sum_{q=1}^J \mathbf{m}_q +\mathbf{r}
\end{equation}
\begin{figure}[t!]
\centering
\subfigure[]{\adjincludegraphics[trim={0 20 0 33},clip,scale=0.4]{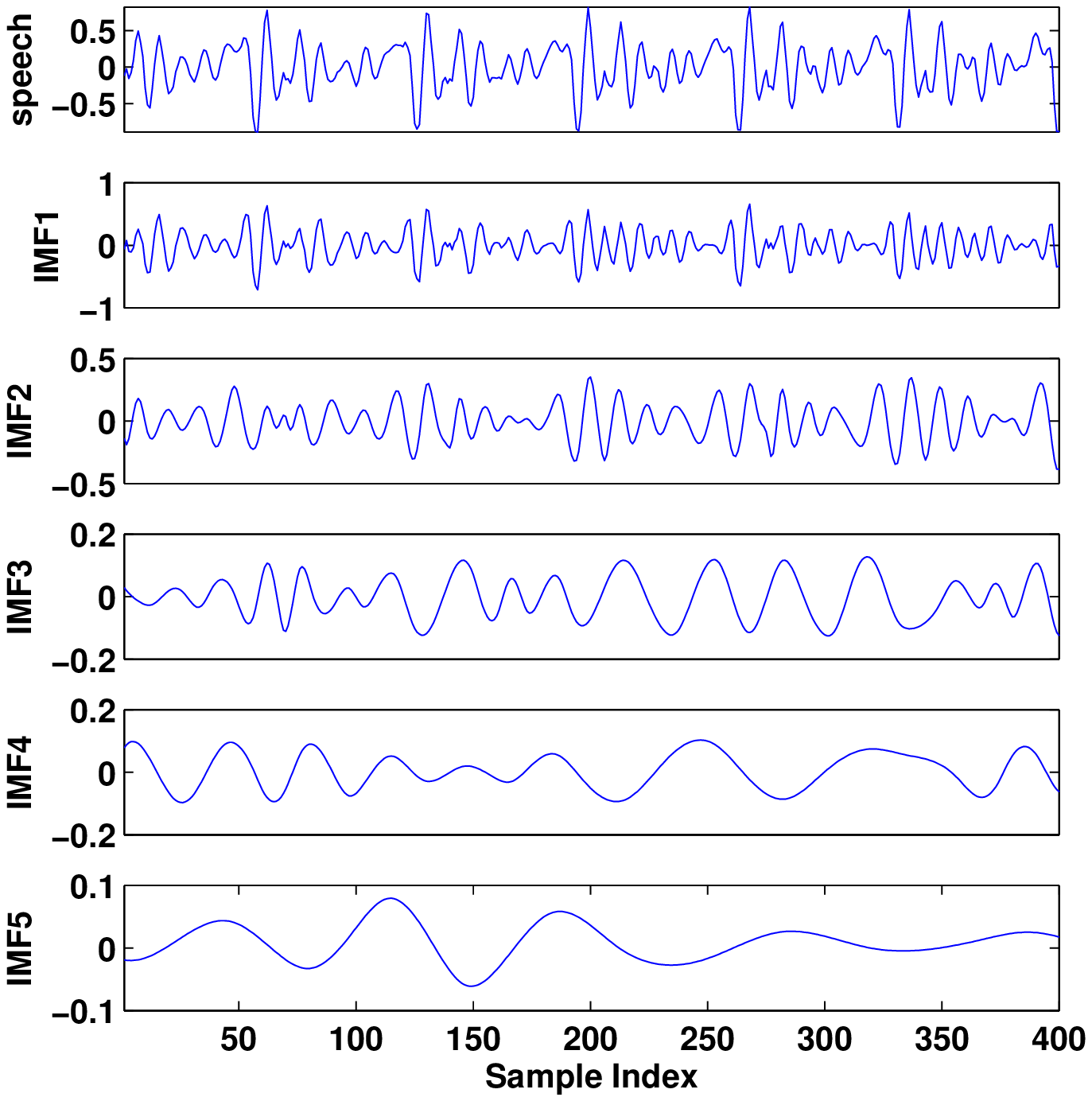}}
\subfigure[]{\adjincludegraphics[trim={0 20 0 25},clip,scale=0.4]{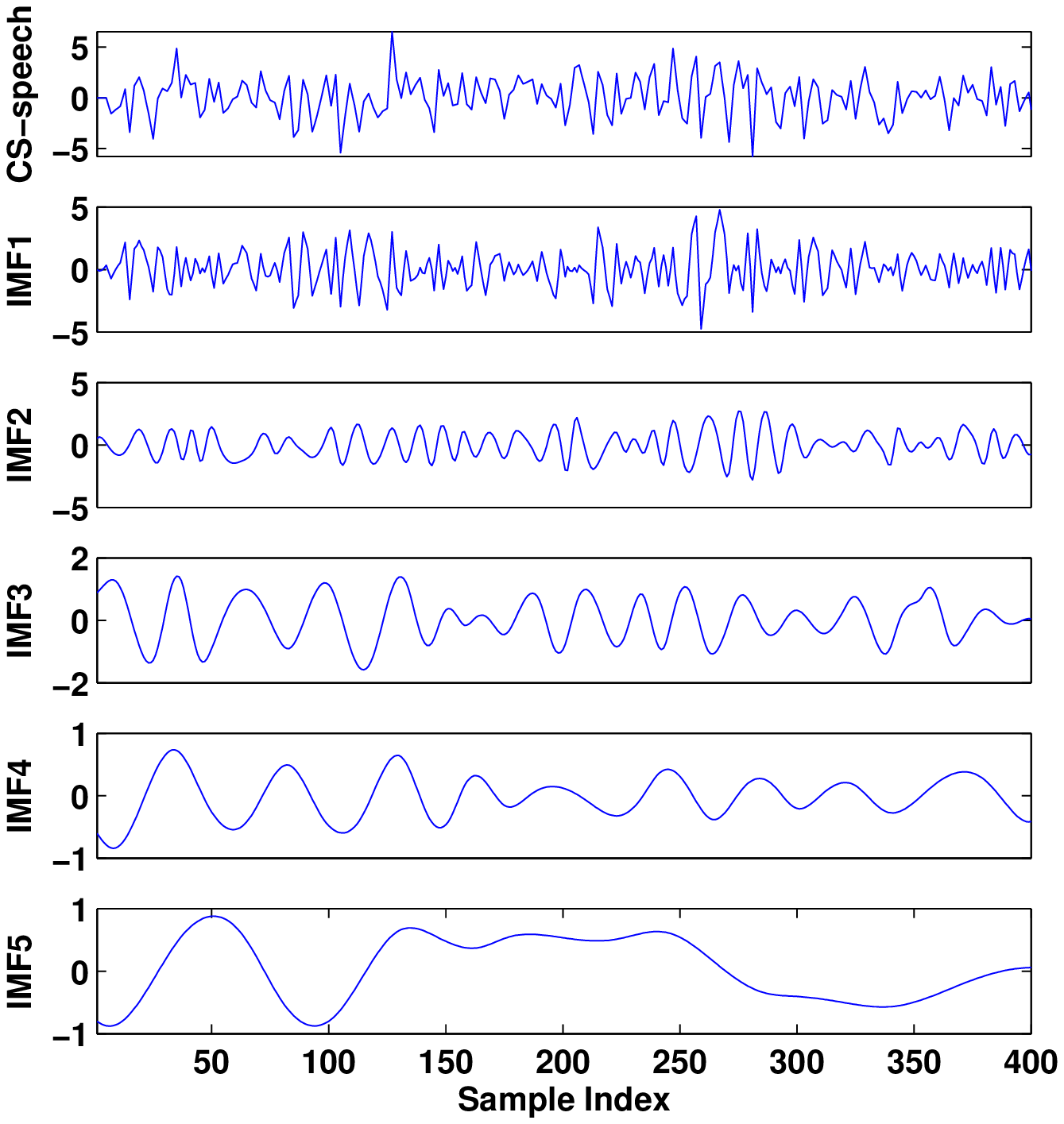}}
\caption{\small EMD decomposition of a voiced frame of (a) orignal speech signal and (b) interpolated compressive speech signal}
\label{fig5}
\end{figure} 
i.e., a sum of $J$ orthogonal modes $\mathbf{m}_q \in \mathbf{R}^m$ and a residual trend $\mathbf{r} \in \mathbf{R}^m$ \cite{ES118}. In order to achieve efficient decomposition, our approach uses the modified EMD algorithms called the Ensemble Empirical Mode Decomposition (EEMD) as proposed in \cite{ES120}. Figs. \ref{fig2} and \ref{fig5}(a) shows an example of compressive and corresponding original voiced speech frame along with the first $5$ extracted IMFs respectively. One can observe that most of the IMFs extracted using compressive samples (Fig. \ref{fig2}) show similar behavior as in case of the IMFs extracted using raw speech samples (Fig. \ref{fig5}(a)). Thus, one can use these IMFs directly to build the dictionary. The extracted IMFs being orthogonal results in a dictionary having  good coherence bounds. Further, the biggest advantage of the proposed approach is its time complexity, which follows from the fact that extracting IMFs and building the dictionary does not require the sensing matrix to be known. However, there are still two major issues in building the dictionary in order to recover the signal: (1) dimensionality of dictionary atoms, and (2) building a dictionary of appropriate size.  

\subsection{Dimensionality of dictionary atoms }
In order to recover the original speech frame, the dimensionality of each dictionary atom should be equal to that of the speech frame (see Eq. (\ref{eq1})). However, any extracted IMFs from the CS measurement vector will have low dimensionality. Further, low sampling rates also affects the performance of EMD. It has been shown that EMD can still be effective (within tolerable limits) if the signal is interpolated such as by Fourier and cosine interpolation methods.  Hence, we used the raised cosine EEMD method \cite{ES121} (with roll-off factor $\beta=1$) to extract IMFs of appropriate dimensions. As an illustration, we have plotted the extracted IMFs of the compressive speech frame considered in Fig. \ref{fig2} after interpolation using EEMD in Fig. \ref{fig5} (b). It can be observe that the IMFs are now more structured as in case of the original speech signal, and can help in learning a better dictionary. 

\subsection{Dictionary size}
\label{CSEMD_1}
Speech signal is generally processed on short frame basis, and  a dictionary build using extracted $J$ IMFs for each compressive speech frame will make it highly overcomplete. However, note that an IMF at each level of decomposition has different scale and structural information. Hence, to restrict the atoms in the dictionary to a desired number, the extracted IMFs from the $J^{th}$ level across all frames are clustered using K-means algorithm. Now the cluster centers are used as dictionary atoms and the number of clusters depends on number of atoms we wish the dictionary to have from each level. To have a sparser representation, more atoms should come from initial levels which contains more structures/patterns as compared to other levels. Algorithm \ref{t2} shows the pseudo-code of the proposed approach.

Note that apart from the presented approach, one is free to explore any variation of EMD algorithm, clustering approach or some other optimal way to build the dictionary from the extracted IMFs. Also, apart from batch processing on all compressive frames, the dictionary can be learned on-line,  where the dictionary atoms are updated as soon as a new frame is available for processing.

\subsection{Computational Complexity}
The time complexity of EMD for extracting all possible IMFs from $L$ $n$-dimensional signal frames approximately scales to $\mathcal{O}(nL\log n)$, that is equal to that of Fast Fourier transform. Further, the complexity of clustering using K-means algorithm is approximately $\mathcal{O}(nLKi)$, where $K$ is the number of clusters and $i$ the number of iterations until convergence. Thus, the overall complexity of the proposed approach is less as compared to conventional DL methods, for which the time complexity per iteration scales to $\mathcal{O}(n^2L)$, and in some cases to $\mathcal{O}(n^3L)$ \cite{ES17}.

\begin{algorithm}[h!]
\centering \footnotesize
\caption{\footnotesize CS-EMD algorithm}
\setlength\extrarowheight{3pt}
\noindent
\begin{flushleft}
\textbf{Inputs:} Compressive signal matrix $\mathbf{Y}=[\mathbf{y}_1\ldots\mathbf{y}_L]$, and sensing matrix $\mathbf{\Phi}$   \\ 
\textbf{Outputs:} Recovered signal matrix $\mathbf{X}=[\mathbf{x}_1\ldots\mathbf{x}_L]$\\
\textbf{Initialization:} $\mathbf{\Psi}=[\cdot]$, $J$, $\epsilon$, $\beta$ and  $K_q \;\forall_q$ s.t. $\displaystyle d=\sum_q K_q$
\end{flushleft}
\begin{algorithmic}[1]
\vspace{-1.8em}
\Statex \textbf{Preprocessing Stage}
\State  Compute $\mathbf{Y}^{'}=[\mathbf{y}_1^{'}\ldots\mathbf{y}_L^{'}]$, using cosine interpolation on $\mathbf{Y}$
\Statex \textbf{Dictionary Learning stage}
\Statex \hspace{.3em} \textbf{for: $i=1\; \mathrm{to}\; L$}
\State Compute $J$ IMFs $\mathbf{m}_{qi}, q=1\ldots J$ from $\mathbf{y}_i^{'}$ using EMD
\Statex \hspace{.3em} \textbf{end for}
\Statex \hspace{.3em} \textbf{for: $q=1\; \mathrm{to}\; J$}
\State Collect $q^{th}$ IMFs $\mathbf{m}_{qi}, i=1\ldots L$ as a column of matrix $\mathbf{M}_q$
\State Cluster columns of matrix $\mathbf{M}_q$ into $K_q$ clusters 
\State Collect cluster centroids as columns of matrix $\mathbf{C}_K$
\State Update Dictionary using cluster centroids as $\mathbf{\Psi}=[\mathbf{\Psi}\; |\; \mathbf{C}_K]$
\Statex \hspace{.3em} \textbf{end for}

\Statex \hspace{.3em} \textbf{Sparse Coding and Signal Recovery stage}
\State Solve $\mathbf{\hat{A}}=\argmin \; \Vert\mathbf{A}\Vert_1\; \mathrm{s.t.} \;   \Vert \mathbf{Y}-\mathbf{\Phi\Psi}\mathbf{A}\Vert_F^2=\Vert \mathbf{Y}-\mathbf{D}\mathbf{A}\Vert_F^2 < \epsilon$
\State Recover signal matrix as $\mathbf{\hat{X}}\approx\mathbf{\Psi\hat{A}}$
\end{algorithmic}
\label{t2}
\end{algorithm}

\section{Experimental Results}
\label{EXP}
In each experiment, speech is processed on a short time frame basis, where framing is achieved by applying a $50$ ms long Hanning window with the frame overlap set to $50\%$. The sensing matrix $\mathbf{\Phi}$ is chosen to be a random Gaussian matrix with a compression ratio $m/n = 0.5$ unless otherwise stated. The maximum number of IMFs extracted using EEMD (with noise realizations $N_e=50$) for each compressive speech frame is set to $5$. A dictionary containing $600$ atoms is learned for each speech utterance (sampled at $8$ KHz) taken from KED TIMIT corpus \cite{ES66}. As initial IMF levels contribute more towards the overall signal approximation the number of dictionary atoms chosen empirically from each IMF level across all frames after clustering are $140$, $140$, $110$, $110$, and $100$ respectively. We conducted experiments on a Quad-Core Intel i$7$ machine at $3.5$ GHz, $12$ Gb RAM, using MATLAB and under Win$8$ operating system. For reasons of brevity, we shall focus on signal recovery, but the proposed dictionary can be readily applied to other speech applications also.

\subsection{Speech recovery from compressive measurements}
In this experiment we assumed that only compressive measurements of a speech utterance are available at the decoder. We considered multiple speech utterances, and for each one a dictionary is learned using the method presented in Section \ref{CSEMD}. The learned dictionary is then applied in CS framework to obtain the sparse representation of each speech frame using $l_1$-minimization, solved using YALL1 package \cite{ES132}. The speech utterance was then reconstructed using standard overlap and add method. 

\begin{figure}[t!]
\centering
\adjincludegraphics[trim={0 0 0 0},clip,scale=0.4]{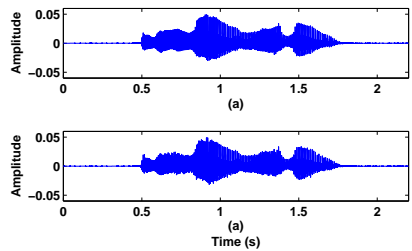}
\caption{(a) Original speech signal. (b) Recovered speech signal from compressed measurements at compression ratio $m/n$ of  $0.5$.}
\label{fig3}
\end{figure}

\begin{figure}[t!]
\centering
\adjincludegraphics[trim={0 0 0 0},clip,width=6cm,height=4.5cm]{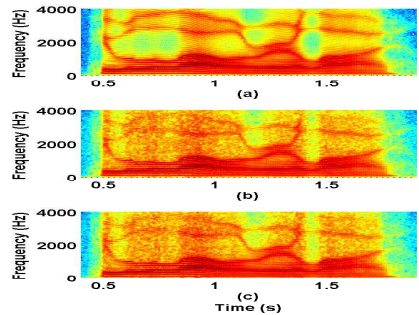}
\caption{Spectrogram of (a) original speech signal; (b) and (c) recovered speech signal from compressed measurements at compression ratio $m/n$ of  $0.5$ and $0.7$ respectively.}
\label{fig4}
\end{figure}

Figure \ref{fig3} shows an example of the original and the reconstructed speech waveform, along with spectrogram plots shown in Figure \ref{fig4}. One can observe that the proposed method is able to recover the speech signal well. However, as observed in Fig. \ref{fig3}(b), the first few extracted IMFs are generally corrupted, and as a results the higher frequency bands of the recovered speech are also distorted. This is also supported by a lower perceptual evaluation of speech quality (PESQ) score for the recovered speech using the proposed approach, compared to other recovery based DL methods as shown in Table \ref{t1}. However, some reduction in speech quality is acceptable, considering the time complexity gain achieved via the proposed approach. To illustrate this, Table \ref{t1} also show the average CPU run times to recover a speech utterance of approximately $3$ sec (including the time for dictionary learning), and the results confirms that the proposed approach is indeed fast compared to existing approaches. Note that for the proposed approach run time is dominated by sparse coding stage.

\subsubsection{Discussion}
Our experiments shows that one can recover a speech signal directly from compressive samples, provided the CS measurements preserve structural properties of the speech signal. The choice of sensing matrix is crucial and if a sensing matrix is carefully chosen or designed one can improve the performance of the proposed approach by learning a better dictionary. In fact, compared to random matrices such as Gaussian/Bernoulli matrices, the performance of the proposed approach increases (as shown in Table \ref{t1}), in case of efficiently designed matrices such as sparse Gaussian and structurally random matrices\footnote{We observed only marginal improvement when sensing matrix other than Gaussian was employed with existing approaches.}. All such matrices do preserve the envelope but fails to preserve the pitch related speech variations in the extracted IMFs, and hence they  result in poor recovery as compared to other recovery based methods. Note that our goal is to recover speech signals from CS measurements at the decoder having limited resources both in terms of storage and computation.

In fact, the extracted IMFs can reveal important properties about speech segments. Hence, the proposed approach is also promising in various inference problems where actual signal recovery is not required, and only CS samples (which require less memory) are available  e.g., voiced/nonvoiced speech detection \cite{ES144}. In such cases, there is even no need to know anything about the sensing matrix used to acquire the signal. However, we defer this or any other extensions to future work. 

\section{Summary}
\label{sec:summary}
In this paper, we have proposed a fast reconstruction free DL approach for compressive speech signals. We show that it is indeed possible to learn a dictionary using only compressive speech samples, and hence the proposed approach is promising in resource-constrained environments. EMD decompositions of compressive speech samples are used to form the atoms of the dictionary, and is motivated by the fact that CS samples have envelop similar to the envelop of original speech samples. Preliminary result on signal recovery experiment, show that the proposed approach can be an alternative to the existing explicit and implicit CS recovery methods. The full potential of this new approach is yet to be realized, and additional work is required to establish the gains. In our future research, we wish to extend this approach to some inference problems, where actual signal recovery is not required. One possible extension is to incorporate the proposed approach in various speech applications such as voice activity detection, speaker identification or speech recognition.

\begin{table}[t]
\centering 
\caption{\footnotesize Comparitive Analysis of Different Methods for Signal Recovery averaged for $20$ utterances over $10$ trials.}
\vspace{1em}
\setlength\extrarowheight{2pt}
\noindent
\footnotesize
\begin{tabular}{|l|l|c|c|p{3.2em}|}
\hline
\textbf{Method}    & \textbf{CS Matrix} & \textbf{DL Iterations} & \textbf{PESQ} & \textbf{Runtime}       \\ \hline
\multirow{4}{*}{CS-EMD} & Sparse-Gaussian & \multirow{4}{*}{N.A} &  2.92  & \multirow{4}{*}{0.83 min} \\ \cline{2-2} \cline{4-4}
            &    SRM \cite{ES142}     &  &  2.91  &                   \\ \cline{2-2} \cline{4-4}
            &  Gaussian               &  &  2.90  &                   \\ \cline{2-2} \cline{4-4}
            &  Bernoulli \cite{ES143} &  &  2.84  &            \\ \hline
Blind CS    &  Gaussian  & 20 &   2.97   &  5 min                 \\ \hline
IHT         &  Gaussian  & 20 &   3.10   &  3 min                 \\ \hline
\end{tabular}
\label{t1}
\end{table}

\bibliographystyle{IEEEbib}
\bibliography{CS-EMD.bib}

\begin{thebibliography}{10}

\bibitem{ES84}
Li~Zeng, Xiongwei Zhang, Liang Chen, Zhangjun Fan, and Yonggang Wang,
\newblock ``Scrambling-based speech encryption via compressed sensing,''
\newblock {\em EURASIP Journal on Advances in Signal Processing}, vol. 2012,
  no. 1, pp. 1--12, December 2012.

\bibitem{ES83}
A.~Asaei, H.~Bourlard, and V.~Cevher,
\newblock ``Model-based compressive sensing for multi-party distant speech
  recognition,''
\newblock in {\em IEEE International Conference on Acoustics, Speech and Signal
  Processing (ICASSP)}, May 2011, pp. 4600--4603.

\bibitem{ES15}
D.~L. Donoho,
\newblock ``Compressed sensing,''
\newblock {\em IEEE Transactions on Information Theory}, vol. 52, no. 4, pp.
  1289--1306, April 2006.

\bibitem{ES18}
I.~Tosic and P.~Frossard,
\newblock ``Dictionary learning,''
\newblock {\em IEEE Signal Processing Magazine}, vol. 28, no. 2, pp. 27--38,
  March 2011.

\bibitem{ES131}
MehmetCenk Sezgin, Bilge Gunsel, and GunesKarabulut Kurt,
\newblock ``Perceptual audio features for emotion detection,''
\newblock {\em EURASIP Journal on Audio, Speech, and Music Processing}, vol.
  2012, no. 1, 2012.

\bibitem{ES17}
Michael Elad,
\newblock {\em Sparse and Redundant Representations - From Theory to
  Applications in Signal and Image Processing.},
\newblock Springer, 2010.

\bibitem{ES129}
R.~Rubinstein, A.M. Bruckstein, and M.~Elad,
\newblock ``Dictionaries for sparse representation modeling,''
\newblock {\em Proceedings of the IEEE}, vol. 98, no. 6, pp. 1045--1057, June
  2010.

\bibitem{ES82}
E.~J. Cand{\'e}s and M.~B. Wakin,
\newblock ``An introduction to compressive sampling,''
\newblock {\em IEEE Signal Processing Magazine}, vol. 25, no. 2, pp. 21--30,
  March 2008.

\bibitem{ES12}
T.~V. Sreenivas and W.~B. Kleijn,
\newblock ``Compressive sensing for sparsely excited speech signals,''
\newblock in {\em IEEE International Conference on Acoustics, Speech and Signal
  Processing (ICASSP)}, April 2009, pp. 4125--4128.

\bibitem{ES19}
D.~Giacobello, M.G. Christensen, M.N. Murthi, S.H. Jensen, and M.~Moonen,
\newblock ``Retrieving sparse patterns using a compressed sensing framework:
  Applications to speech coding based on sparse linear prediction,''
\newblock {\em IEEE Signal Processing Letters}, vol. 17, no. 1, pp. 103--106,
  2010.

\bibitem{ES130}
C.~Studer and R.G. Baraniuk,
\newblock ``Dictionary learning from sparsely corrupted or compressed
  signals,''
\newblock in {\em IEEE International Conference on Acoustics, Speech and Signal
  Processing (ICASSP)}, March 2012, pp. 3341--3344.

\bibitem{ES117}
S.~Gleichman and Y.C. Eldar,
\newblock ``Blind compressed sensing,''
\newblock {\em IEEE Transactions on Information Theory}, vol. 57, no. 10, pp.
  6958--6975, October 2011.

\bibitem{ES28}
Ch.~Srikanth Raj and T.~V. Sreenivas,
\newblock ``Time-varying signal adaptive transform and {IHT} recovery of
  compressive sensed speech,''
\newblock in {\em 12th \mbox{INTERSPEECH}}, August 2011, pp. 73--76.

\bibitem{ES118}
A.~Bouzid and N.~Ellouze,
\newblock ``Empirical mode decomposition of voiced speech signal,''
\newblock in {\em First International Symposium on Control, Communications and
  Signal Processing (ISCCSP).}, March 2004, pp. 603--606.

\bibitem{ES120}
M.E. Torres, M.A. Colominas, G.~Schlotthauer, and P.~Flandrin,
\newblock ``A complete ensemble empirical mode decomposition with adaptive
  noise,''
\newblock in {\em IEEE International Conference on Acoustics, Speech and Signal
  Processing (ICASSP)}, May 2011, pp. 4144--4147.

\bibitem{ES121}
A.~Roy and J.F. Doherty,
\newblock ``Raised cosine filter-based empirical mode decomposition,''
\newblock {\em IET Signal Processing}, vol. 5, no. 2, pp. 121--129, April 2011.

\bibitem{ES66}
``{U}niversity of {E}dinburgh's {KED TIMIT},'' {http://festvox.org/}.

\bibitem{ES132}
J.~Yang Y.~Zhang and W.~Yin,
\newblock ``{YALL1}: Your algorithms for l1,'' http://www.yall1.blogs.rice.edu,
  2011.

\bibitem{ES144}
V.~Abrol, P.~Sharma, and {A.K.} Sao,
\newblock ``Voiced/nonvoiced detection in compressively sensed speech
  signals,''
\newblock {\em Speech Communication}, vol. 72, no. 0, pp. 194 -- 207, 2015.

\bibitem{ES142}
T.T. Do, Lu~Gan, N.H. Nguyen, and T.D. Tran,
\newblock ``Fast and efficient compressive sensing using structurally random
  matrices,''
\newblock {\em IEEE Transactions on Signal Processing}, vol. 60, no. 1, pp.
  139--154, Jan 2012.

\bibitem{ES143}
Gesen Zhang, Shuhong Jiao, Xiaoli Xu, and Lan Wang,
\newblock ``Compressed sensing and reconstruction with bernoulli matrices,''
\newblock in {\em IEEE International Conference on Information and Automation
  (ICIA)}, June 2010, pp. 455--460.

\end{thebibliography}

\end{document}